\documentclass[preprint]{aastex}
\usepackage[]{natbib}
\usepackage{graphics}
\usepackage{emulateapj5}
\usepackage{apjfonts}

\newcommand{\etal}{et al.}  
\newcommand{\per}{\ensuremath{^{-1}}}

\newcommand{\mbh}{\ensuremath{M_\mathrm{\bullet}}}

\newcommand{\kms}{km s\ensuremath{^{-1}}}

\newcommand{\chisq}{\ensuremath{\chi^2}}
\newcommand{\chisqdof}{\ensuremath{\tilde{\chi}^2}}

\newcommand{\msigma}{\ensuremath{\mbh - \sigma}}
\newcommand{\mgb}{\ion{Mg}{1}\emph{b}}

\newcommand{\chandra}{\emph{Chandra}}

\slugcomment{}
\shorttitle{VELOCITY DISPERSION MEASUREMENTS} 
\shortauthors{BARTH ET AL.}

\begin{document} 

\title{A Study of the Direct-Fitting Method \\ for Measurement of Galaxy
  Velocity Dispersions}

\author{Aaron J. Barth\altaffilmark{1,2}, Luis C. Ho\altaffilmark{3},
and Wallace L. W. Sargent\altaffilmark{1}}

\altaffiltext{1}{Palomar Observatory, 105-24 Caltech, Pasadena, CA
91125; barth@astro.caltech.edu}
\altaffiltext{2}{Hubble Fellow}
\altaffiltext{3}{The Observatories of the Carnegie Institution of
Washington, 813 Santa Barbara Street, Pasadena, CA 91101}

\begin{abstract}

We have measured the central stellar velocity dispersions of 33 nearby
spiral and elliptical galaxies, using a straightforward
template-fitting algorithm operating in the pixel domain.  The
spectra, obtained with the Double Spectrograph at Palomar Observatory,
cover both the Ca triplet and the \mgb\ region, and we present a
comparison of the velocity dispersion measurements from these two
spectral regions.  Model fits to the Ca triplet region generally yield
good results with little sensitivity to the choice of template star.  In
contrast, the \mgb\ region is more sensitive to template mismatch and
to details of the fitting procedure such as the order of a polynomial
used to match the continuum shape of the template to the object.  As a
consequence of the correlation of the [Mg/Fe] ratio with velocity
dispersion, it is difficult to obtain a satisfactory model fit to the
\mgb\ lines and the surrounding Fe blends simultaneously, particularly
for giant elliptical galaxies with large velocity dispersions.  We
demonstrate that if the metallicities of the galaxy and template star
are not well matched, then direct template-fitting results are
improved if the \mgb\ lines themselves are excluded from the fit and
the velocity dispersion is determined from the surrounding weaker lines.

\end{abstract}


\section{Introduction}

The recent discovery of a tight correlation between stellar velocity
dispersion and black hole mass \citep[the \msigma\ relation;][]{fm00,
geb00} has placed new emphasis on the importance of accurate velocity
dispersion measurements for the central regions of nearby galaxies.
Since black hole mass is approximately proportional to $\sigma^4$,
even modest errors in $\sigma$ for galaxies with black hole mass
measurements can have a substantial impact on the correlation
\citep{tre02}.  Use of the predictive power of the \msigma\ relation
to obtain estimates of the masses of black holes in galaxy nuclei also
relies on the accuracy of the velocity dispersion measurements.  In
view of these issues, it is worthwhile to examine the level of
agreement between velocity dispersion measurements obtained with
different techniques and from different spectral regions, to determine
the methods that are most likely to yield accurate results.

This paper presents an examination of a simple direct template-fitting
technique operating in the wavelength domain.  Initially, our main
goal was to measure velocity dispersions for the sample of nearby
galaxies observed with \chandra\ by \citet{ho01}, so that estimates of
the black hole masses in these galaxies could be derived by applying
the \msigma\ relation.  We were able to observe most of the galaxies
from this sample during two observing runs.  We also observed some
other nearby galaxies for which no previous velocity dispersion
measurements were available, a few velocity dispersion ``standard''
galaxies from \citet{mce95} for comparison, and several low-redshift
BL Lac objects for which the results have been reported separately
\citep{bhs02a,bhs02b}.  Here, we present velocity dispersions for 33
nearby galaxies and a comparison of measurements obtained from the Ca
triplet and \mgb\ spectral regions.  We also discuss some systematic
issues relevant to the application of the direct template-fitting
method.  In particular, we demonstrate that model fits to the \mgb\
spectral region are sensitive to the [Mg/Fe] abundance ratio, and that
the results are usually improved if the \mgb\ lines themselves are
\emph{excluded} from the fitting region used to determine the velocity
dispersion.

\section{Observations and Reductions}

The data were obtained during two observing runs with the Double
Spectrograph \citep{og82} at the Palomar Hale 5m telescope, in 2001
June and 2002 January.  The red side setup was identical during the
two runs.  For the blue side, we used a 600 lines mm\per\ grating
during the 2001 June run, and a 1200 lines mm\per\ grating during the
2002 January run.  The setups and instrumental dispersions are listed
in Table 1.  Conditions were mediocre during both runs, with typical
seeing of 1\farcs5--2\farcs0 and moderate to thick cloud cover during
most of the nights.  None of the data were obtained in photometric
conditions.

The spectrograph slit was oriented at the parallactic angle, so that
the blue and red spectra would cover the same region of the galaxy.
This allows a more consistent comparison between the red and blue
measurements but with the caveat that our measurements do not
generally correspond to any particular symmetry axis of the galaxy.
Given the mediocre seeing during our observing runs and the fact that
the slit position angle is effectively random relative to the galaxy
major axis, we choose to measure only the bulk velocity dispersion in
a region centered on the nucleus, rather than performing
a detailed study of spatially-resolved kinematics.  The high
signal-to-noise ratio (S/N) of the spectra obtained from these wide
extractions allows us to explore various systematic issues in the
measurement procedure.

Bias subtraction and flat-fielding were performed in the usual manner
using standard IRAF tasks.  Extractions were performed over a width of
3\farcs74, which corresponds to 8 pixels on the red CCD and 6 pixels
on the blue CCD.  This extraction width was chosen so that the blue
and red measurements would cover exactly the same spatial region of
the galaxy, to facilitate comparison of the results.  The spectra were
extracted optimally \citep{hor86} in order to maximize the S/N and
remove cosmic-ray hits.  We verified that there were no differences in
line strength or width between the optimally-weighted spectra and
normally-weighted extractions over the same extraction widths.  The
fractional differences in pixel values between the optimal and normal
extractions were typically at the level of $\lesssim10^{-3}$ for
well-exposed spectra, except at pixels affected by cosmic-ray hits.
Averaged over all galaxies in the sample, the mean S/N per pixel of
the extracted spectra is 107 for the red side data, and 224 and 142
for the blue side data obtained with the 600 and 1200 line gratings,
respectively, in the wavelength regions used to determine $\sigma$.

The spectra were then wavelength calibrated and flux calibrated using
standard techniques.  The deep telluric H$_2$O absorption features
longward of 8900 \AA\ were removed by dividing by the normalized
spectrum of a standard star, but this spectral region is not used in
any of the analysis discussed below.  Finally, a small linear shift
was applied to the wavelength scale of each exposure based on the
wavelengths of night sky emission lines, and in cases where multiple
exposures of the same object were taken, they were averaged together.

\section{Measurements}

\subsection{Method}

Our observations cover the Ca triplet at $\lambda\lambda$8498, 8542,
8662 \AA\ and the \mgb\ triplet at $\lambda\lambda$5167, 5173, 5184
\AA, as well as blends of Fe lines at 5270 and 5335 \AA\ and other
weaker lines in the region surrounding \mgb.  As discussed by
\citet{dre84}, the Ca triplet region is ideal for velocity dispersion
measurements since it is relatively insensitive to differences in
stellar populations.

We performed the measurements of $\sigma$ with a direct
template-fitting routine.  Direct fitting in the pixel domain is the
most straightforward method for measuring velocity dispersions
\citep{vdm94, rix95}.  One of its main advantages is that it is simple
to mask out unwanted spectral features from the fit and compute
\chisq\ only over an exactly specified wavelength range.  This is
particularly useful for galaxies having emission lines, or when
telluric or interstellar absorption lines are in the wavelength range
of interest.  Direct fitting has the disadvantage of being more
computationally expensive than, for example, cross-correlation
\citep{td79} or Fourier quotient \citep{sim74, ssbs77} methods, but
this is no longer a severe limitation with processor speeds currently
available.

To measure $\sigma$, the object and template spectra are rebinned to a
wavelength scale that is linear in $x=\ln \lambda$, so that velocity
shifts can be performed by linear shifts to the log wavelength
scale.\footnote{A redshift $z$ corresponds to a wavelength
transformation $x^\prime = x + \ln[1+z]$; this is closely
approximated by $x^\prime = x + z$ for $z \ll 1$.}  The
logarithmic bin size is chosen to preserve the total number of pixels
in the spectrum.  The object spectrum $O(x)$ is transformed to zero
redshift by setting $O(x_\mathrm{rest}) = O(x_\mathrm{observed} +
\ln[1+z])$, where $z$ is the galaxy's redshift.

Our measurement technique is very similar to the direct-fitting
procedure described by \cite{kel00}.  For simplicity, we assume a
Gaussian profile for the line-of-sight velocity distribution (LOSVD).
Then, the model $M(x)$ is computed as
\begin{equation}
M(x) = \left\{ \left[ T(x) \otimes G(x) \right] +
C(x) \right\} \times P(x),
\end{equation}
where $T$ is the stellar template spectrum, $G$ is the Gaussian
broadening function, $C$ is a featureless continuum, $P$ is a
low-order Legendre polynomial, and $\otimes$ denotes convolution.  The
featureless continuum is a straight line of arbitrary slope that
allows the model to match the dilution of the galaxy spectrum.  A
straight line is sufficient to model the featureless continuum since
our fits are performed only over a small wavelength range.  The
multiplicative polynomial $P$ allows for possible differences in
continuum shape or reddening between the object and template without
requiring an explicit determination of the reddening.  The order of
the polynomial is kept low to ensure that it does not affect the
widths of individual absorption features; we discuss this issue in
more detail below.  To determine the goodness of fit we calculate
$\chisq$ according to
\begin{equation}
\chisq = \sum_i \left( \frac{M_i - O_i}{\epsilon_i} \right)^2,
\end{equation}
where $\epsilon(x)$ is the error spectrum corresponding to the object
spectrum $O(x)$, and the calculation is performed over all pixels $i$
in the fitting region.  The template spectra are assumed to be
noise-free as they have much higher S/N than the galaxy spectra.  The
fit is optimized by minimizing \chisq\ using a downhill simplex
algorithm \citep{nr}.  To ensure that the global minimum of \chisq\ is
found, the fit is repeated several times with different search
parameters.   This method can be generalized to include
higher-order moments of the velocity profile \citep{vdm94} or a blend
of templates of different spectral types \citep{rw92, rix95}, but in
general we found that good fits to the Ca triplet region could be
achieved using the simple model of Gaussian-broadened K-giant spectra,
at least over the fairly large spectroscopic aperture employed in this
study.

For the pixel errors $\epsilon_i$ we used the error spectra returned
by the IRAF optimal extraction algorithm, calibrated in the same
manner as the galaxy spectra.  For well-exposed spectra, these errors
were found to be essentially identical to error spectra calculated
from the Poisson noise, background noise, and CCD readout noise for a
given extraction.  The uncertainty on $\sigma$ was calculated by
finding how much $\sigma$ must be displaced from its best-fit value,
with all other parameters allowed to float freely, in order to
increase \chisq\ by unity relative to its minimum value.  

Prior to calculating the uncertainty, the pixel errors $\epsilon_i$
were rescaled so as to yield $\chisqdof = 1$ (where \chisqdof\ denotes
\chisq\ per degree of freedom).  This is conservative in that it
increases the size of the error bars on $\sigma$, but in practice it
results in only a small increase in the uncertainty estimates.  We
note that in some cases, particularly for observations with lower than
average S/N on the red side, the best fit obtained using the original
error spectrum would sometimes yield $\chisqdof < 1$; this suggests
that the optimal algorithm in IRAF tends to overestimate the pixel
errors by a small amount, $\sim10-20\%$, at low S/N.  When this
occurred, the pixel errors were not rescaled.  We also performed tests
of our fitting routine using equal weighting for all pixels rather
than using the error spectrum to calculate \chisq.  Equal weighting
gave results virtually identical to weighting by the pixel errors,
with differences of $\leq1$ \kms\ in both $\sigma$ and its
uncertainty. Thus, measurements obtained by the direct fitting method
are not adversely affected if no error spectrum is available.

For each measurement, we used 8 template stars of spectral type
G7--K5, with luminosity class III or IV; the stars are listed in Table
2.  The uncertainty in $\sigma$ is dominated by template mismatch in
many cases, particularly for the blue side data, as the scatter in
$\sigma$ values for different templates exceeds the uncertainty in a
single fit.  To account for this scatter, the final measurement
uncertainty given for each galaxy is the sum in quadrature
of the fitting uncertainty for the best-fitting template and the
standard deviation of the measurements from all 8 template stars.

\subsection{Fitting Region}

\begin{figure}
\begin{center}
\plotone{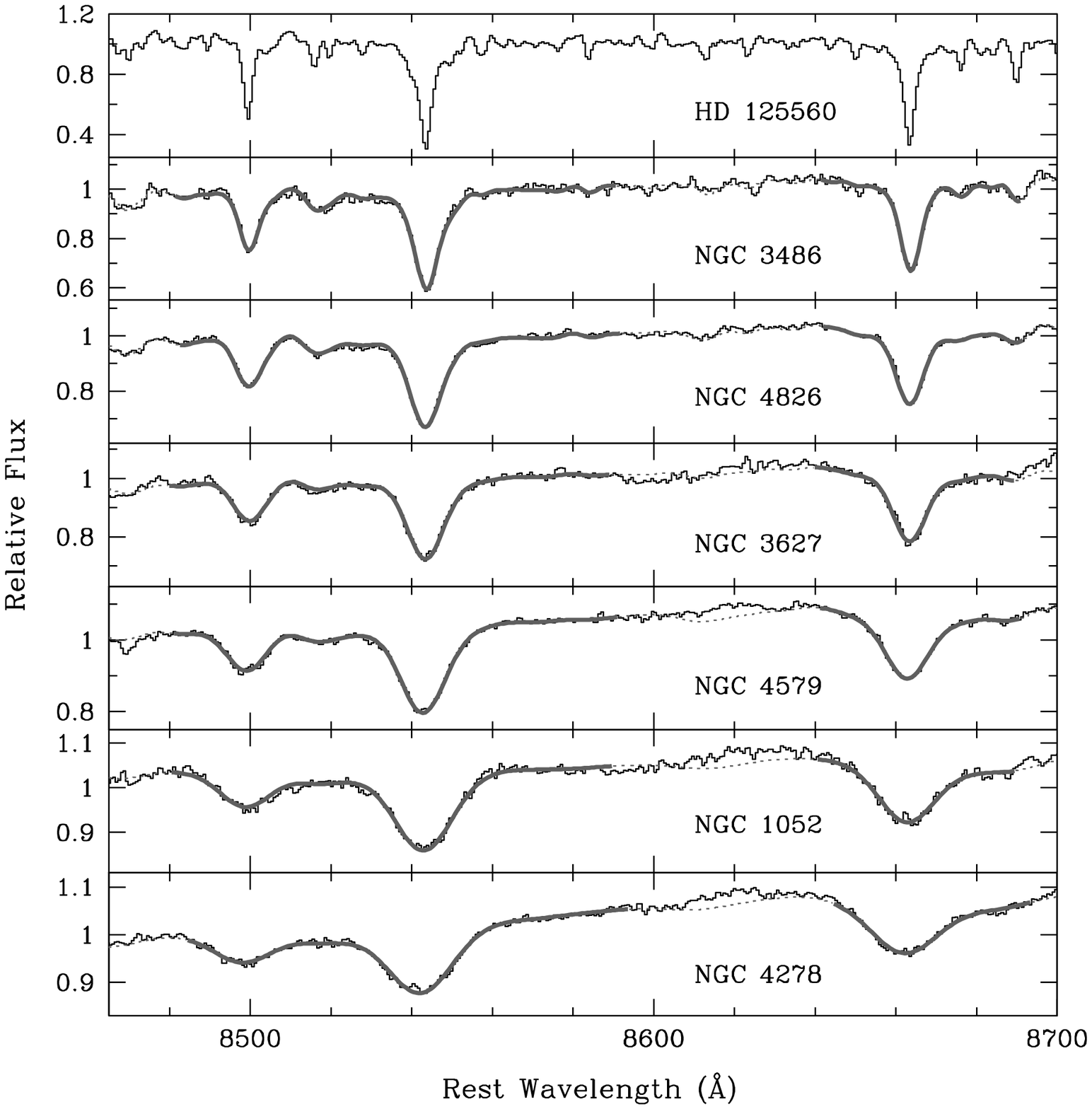}
\end{center}
\caption{Best fits of broadened stellar templates to the spectra of
  velocity dispersion standard galaxies from the \citet{mce95}
  catalog.  Regions used to compute \chisq\ are denoted by a solid
  thick line, and regions excluded from the fit are indicated by a
  dotted line.The spectrum of the K3 III star HD 125560 is shown in
  the top panel.
\label{rfits}}
\end{figure}

The fitting regions were chosen to include portions of the spectrum
that are sensitive to $\sigma$ but not contaminated by emission
lines.  On the red side, we used a wavelength region closely
surrounding the Ca triplet lines, $8480-8690$ \AA.  The continuum
region $8590-8640$ \AA\ was exluded from the fit as the fits are often
poor over this region and there may be [\ion{Fe}{2}] emission at 8617
\AA\ in some galaxies \citep{vdm94}.  The continuum regions near the
Ca triplet lines have very little sensitivity to $\sigma$ in any case.
Figure \ref{rfits} illustrates the quality of the red side fits over a
wide range in $\sigma$.  The exclusion of the region $8590-8640$ \AA\
from the fitting region is clearly justified by the poor template
match over this wavelength range for NGC 4579, NGC 1052, and NGC 4278.

The blue side data presented additional challenges.  Our original
intent was to compare the Ca triplet measurements with velocity
dispersions measured from the \mgb\ region, and after some trial and
error we chose the region $5040-5430$ \AA\ over which reasonable fits
could usually be obtained.  However, the template spectra often give a
very poor match to the \mgb\ lines themselves.  While the surrounding
region of continuum and weak absorption lines could generally be fit
adequately, we found that the fits were usually dramatically improved
(in the sense of having a lower value of \chisqdof) by excluding \mgb\
from the fit.  The change in \chisqdof\ is particularly dramatic for
galaxies with high velocity dispersions.  We also found it necessary
to exclude a small region around 5200 \AA, as many galaxies have weak
[\ion{N}{1}] $\lambda5200$ \AA\ emission.  Even when this line is not
clearly visible in the spectra, the fits often reveal a slight excess
in the galaxy spectra near 5200 \AA\ relative to the broadened
templates.  For our final fits, we excluded the range $5150-5210$ \AA\
from the calculation of \chisq.

\begin{figure}
\begin{center}
\plotone{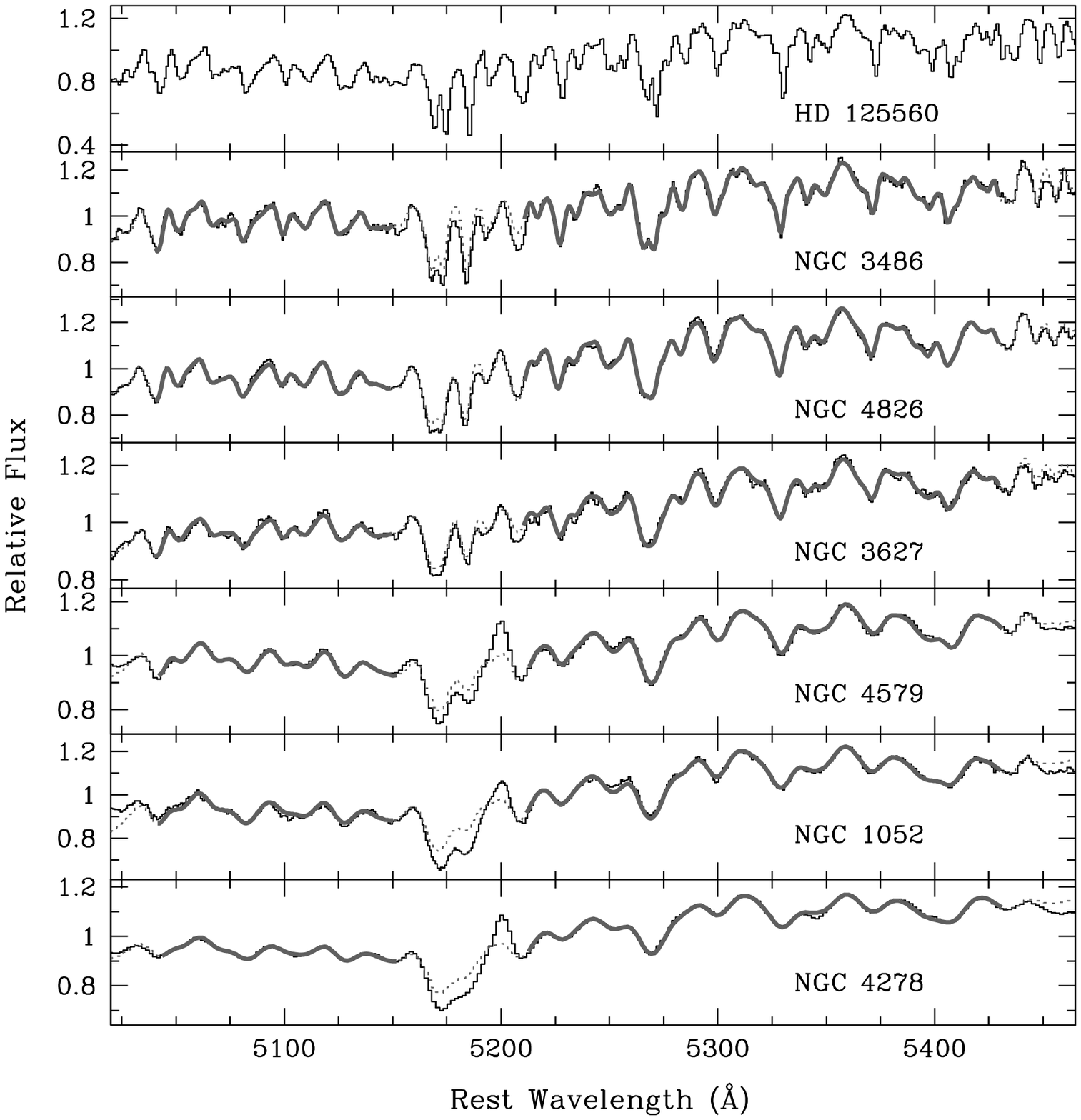}
\end{center}
\caption{Best fits of broadened stellar templates to the spectra of
  velocity dispersion standard galaxies from the \citet{mce95}
  catalog.  Regions used to compute \chisq\ are denoted by a solid
  thick line.  Regions excluded from the fit (the \mgb\ triplet and [N
  I] $\lambda5200$) are indicated by a dotted line.  The spectrum
  of the K3 III star HD 125560 is shown in the top panel.
\label{bfits}}
\end{figure}

The difficulty of fitting the \mgb\ profile is a consequence of the
correlation between \mgb\ line strength and $\sigma$ \citep{ter81,
dre87}.  More specifically, since the ratio [Mg/Fe] in elliptical
galaxies is correlated with velocity dispersion \citep{wfg92, tra98,
kun01}, galaxies with high $\sigma$ will have stronger \mgb\
absorption relative to the surrounding Fe lines, in comparison with
typical nearby giant stars used as velocity templates.  Our template
stars were chosen based on their spectral type and magnitude, without
regard to metallicity, and most have slightly subsolar to slightly
above-solar [Fe/H].  The fits shown in Figure \ref{bfits} demonstrate
that with these template stars, it is not possible to achieve a good
fit to both the \mgb\ lines and the surrounding region (dominated by
Fe blends) for galaxies of high $\sigma$.  

As a comparison, we attempted to perform fits using the \mgb\ lines
alone, over the range 5150--5190 \AA.  However, since the Mg lines
cover such a small wavelength range, the fits were very poorly
constrained even after restricting both the multiplicative polynomial
and the featureless continuum to zeroth order (i.e., both flat in
$f_\lambda$).  Consequently, no useful results could be obtained from
fitting models only to the \mgb\ lines.

In principle, the [Mg/Fe] mismatch problem could also be alleviated to
some extent by observing template stars having a range of super-solar
metallicities.  However, it would be difficult to find an exact
stellar match to the [Mg/Fe] ratio of any given galaxy spectrum, and
it would be very time-consuming to observe a grid of template stars
with a range in both spectral type and metallicity.
 
We also note that both the \mgb\ and the Ca triplet lines are
intrinsically strong features that are subject to pressure
broadening. The widths of these lines in a galaxy spectrum are due to
a composite population of stars with a range of surface gravities and
therefore a range in intrinsic linewidths.  For galaxies with very
small velocity dispersions, or for objects such as young super star
clusters, it may be preferable to measure velocity dispersions only
from regions containing weaker, intrinsically narrow lines, such as
the region just redward of \mgb\ \citep[e.g.,][]{hf96}.

\subsection{Tests of the Measurement Routine}

We performed tests to determine whether the measurement routine would
yield correct results on a galaxy of known velocity dispersion,
including some degree of template mismatch.  To perform the tests, we
began with a composite template spectrum made by combining a K2 III
star, a G2 V star, and a featureless continuum.  The weights were
chosen so that the total flux in the composite spectrum was 40\% K2
III, 40\% G2 V, and 20\% featureless continuum.  The spectrum was then
broadened by convolution with Gaussians having $\sigma = 50-300$ \kms,
and its spectral shape was adjusted by multiplication by a 4th-order
Legendre polynomial.  The polynomial coefficients were restricted so
that the continuum flux at any wavelength was modified by $<5\%$.
Finally, Poisson noise was added to the broadened spectrum to give S/N
= 120 per pixel.  The velocity dispersions were measured using the
same techniques described above and the same set of template spectra.
For the blue side, the simulated data were created using spectra
observed with the 1200 line grating.

For both the blue and red sides, the measurement routine is very
successful at recovering the correct velocity dispersion over the
entire range of $50-300$ \kms.  In all cases, the measured dispersion
is within 6 \kms\ of the input dispersion.  The red side measurements
are systematically closer to the input dispersion than the blue side
results are, however: the RMS disagreement between the measured and
input velocity dispersions is 3.0 \kms\ for the red side and 4.4 \kms\
for the blue side.  This is not by any means a complete test of the
method; a full examination would require testing the measurement
routine as a function of velocity dispersion, S/N, and degree of
template mismatch, for both the red and blue spectral regions.
Nevertheless, it does demonstrate the accuracy of the method for a
somewhat simplified input model.  As expected, the results were less
accurate for velocity dispersions smaller than the instrumental
dispersion.  On the blue side, for an input model with
$\sigma_{\mathrm{in}}=25$ \kms, the measurement routine gave $\sigma =
34 \pm 5$ \kms.  The red side, with an instrumental dispersion of 25
\kms, returned $\sigma=25\pm3$ \kms\ for an input dispersion of 25
\kms, but for input dispersions of $\lesssim15$ \kms\ it was not able
to yield useful results.

\begin{figure}
\plotone{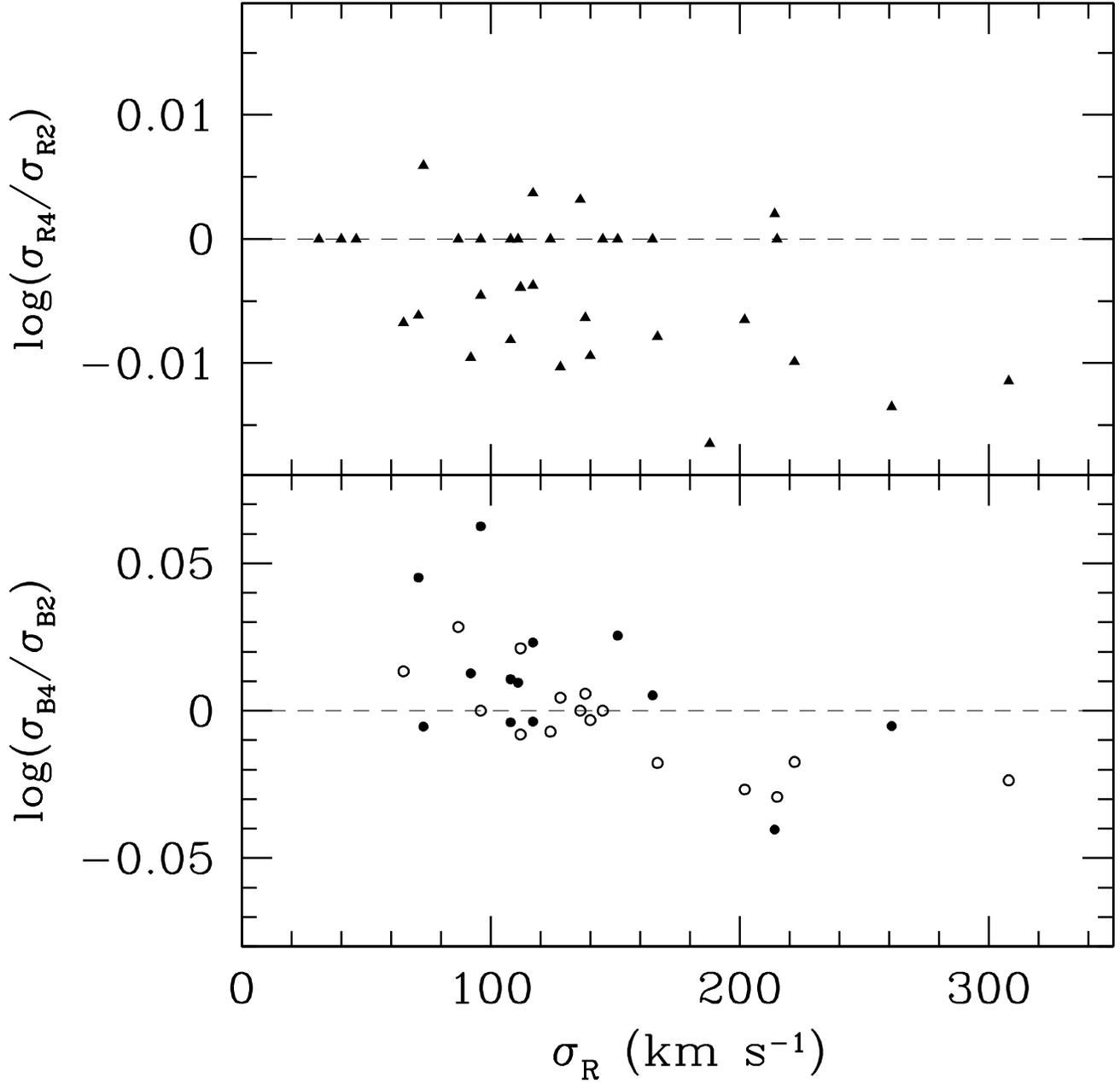}
\caption{Comparison of measurements performed using a multiplicative
  polynomial of order 2 or of order 4, for the red (upper panel) and
  blue (lower panel) spectra. Note the different vertical scales for
  the upper and lower panels.  For the blue side measurements, the
  filled and open circles represent galaxies observed with the 600 and
  1200 line gratings, respectively.
\label{order}}
\end{figure}

We investigated one other aspect of the measurement routine, the order
of the multiplicative polynomial used to adjust the template's
continuum shape to match the galaxy.  Our final measurements were
performed using a quadratic polynomial on both the red and blue sides.
Figure \ref{order} demonstrates how these results change if a 4th
order polynomial is used.  The red side results are nearly unchanged.
However, the blue side measurements with the two polynomial models
disagree by a larger amount, with an RMS difference of 5\% for our
sample.

There is no \emph{a priori} reason to prefer any particular polynomial
order, as long as the order is low enough not to introduce any
structure on scales close to the width of individual or blended
spectral features.  Thus, we interpret the variation of $\sigma_B$
with polynomial order as a source of systematic uncertainty in the
blue measurements; it is an additional reason to prefer the Ca triplet
region for measurement of $\sigma$.  For high values of $\sigma$, it
appears that using a 4th-order polynomial systematically reduces the
measured value of $\sigma$; this could be a consequence of the
polynomial function over-fitting the shallow, blended, broad
absorption features in the galaxy spectrum.  However, there does not
appear to be an obvious explanation for the systematic rise in the
derived value of $\sigma$ when using a higher-order polynomial fit to
galaxies with $\sigma \lesssim 175$ \kms.  \citet{kel00} discuss the
issue of the polynomial order in direct-fitting routines; they find
that their results are not very sensitive to the polynomial order,
with $<1\%$ differences in velocity dispersions derived with
polynomial orders of 4, 5, or 6.  The higher sensitivity we find may
be due to the fact that our measurements are performed over a small
wavelength region, so that higher-order polynomials may begin to fit
individual absorption features in addition to the overal spectral
shape.  Also, our fits do not require such high-order polynomials
since our spectra are flux-calibrated and the galaxy redshifts are
small, minimizing any instrumental differences in spectral shape
between the galaxies and template stars.

Since the red side is dominated by just a few strong lines surrounded
by a relatively simple continuum, it is less susceptible to mismatch
in the continuum shape.  For individual galaxies the agreement between
blue and red measurements can be either better or worse when using the
higher-order polynomial on the blue side.  Considering the entire
sample, the RMS difference between blue and red measurements is
virtually identical whether the blue side polynomial order is 2, 3, or
4, and we somewhat arbitrarily choose the quadratic model for the
final blue side measurements.  This issue may have a similar effect on
velocity dispersion measurements obtained with other techniques, since
other methods also require normalization of the continuum shape of the
galaxy and template star, and this normalization is usually carried
out by multiplication by a polynomial \citep[e.g.,][]{fih89, dal91}.

\section{Results and Discussion}

\subsection{Comparison of blue and red side results}

Table 3 lists the velocity dispersions measured from the blue and red
side spectra.  For three galaxies (NGC 404, NGC 660, and NGC 6503),
the velocity dispersion was too small to be measured from the blue
side data.  We did not attempt to fit models to the blue spectrum of
Arp 102B as it contains a number of weak emission lines in this
region.  In general, we consider the red side results to be more
reliable for the reasons described above.  One simple measure of the
relative degree of template mismatch between the red and blue sides is
the goodness of fit as measured by \chisqdof.  The mean value of
\chisqdof\ for all galaxies is 1.12 on the red side, and 3.03 on the
blue side.  Given that the same LOSVD model was used for the red and
blue side data, this difference in \chisqdof\ clearly demonstrates the
larger degree of template mismatch on the blue side.  In addition to
template mismatch issues, the red measurements are aided by higher
spectral resolution.  The low spectral resolution of the blue side
data obtained with the 600 line grating is clearly reflected in the
large error bars for these measurements, in comparison with the data
obtained with the 1200 line grating.

\begin{figure}
\plotone{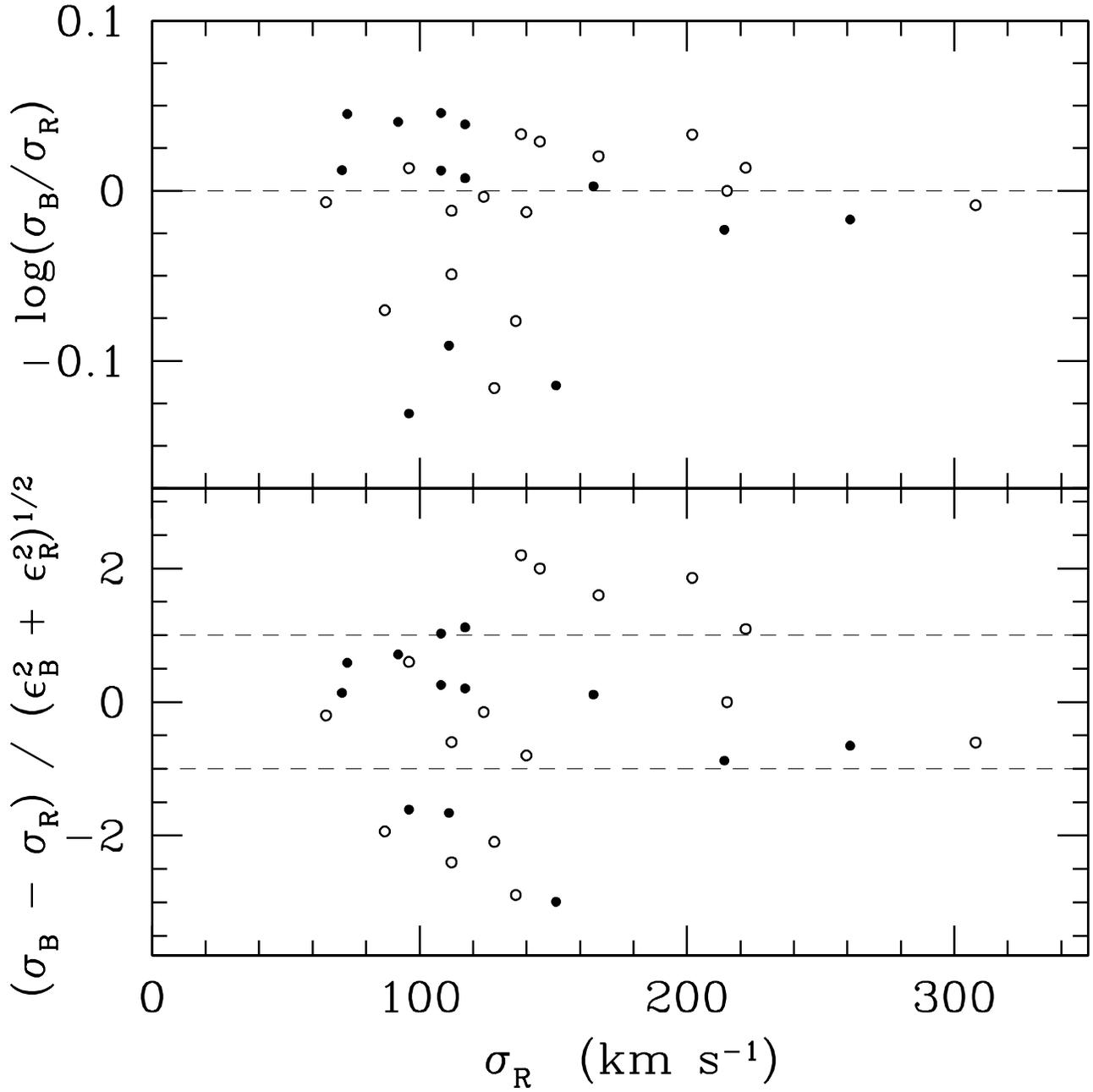}
\caption{Comparison of blue and red side measurements.  \emph{Upper
  panel:} Logarithm of $\sigma_B/\sigma_R$.  \emph{Lower panel:} The
  difference between $\sigma_B$ and $\sigma_R$, normalized to the
  measurement uncertainties $\epsilon$.  Filled circles and open
  circles represent galaxies observed with the 600 and 1200 line
  gratings, respectively. 
\label{brcompare}}
\end{figure}

To determine whether the red and blue results are in agreement within
their uncertainties, we compute the statistic $\delta = (\sigma_B -
\sigma_R) / (\epsilon_B^2 + \epsilon_R^2)^{1/2}$.  If $|\delta| \leq
1$, then the difference between the red and blue measurements is
consistent with zero, and the two measurements are considered to be in
agreement.  Figure \ref{brcompare} shows the results of this test.
Out of 29 galaxies with blue and red measurements, 15 (or 52\%) agree
within the estimated $1\sigma$ uncertainties.  The worst disagreements
are at the $3\sigma$ level.  This suggests that the measurement
uncertainties may be somewhat underestimated, particularly for the
blue side data.  

The systematic uncertainty in the choice of the polynomial order for
the blue measurements may be largely to blame for this situation.  As
described above, there appears to be an additional uncertainty of
roughly 5\% in the blue side measurements due to the choice of
polynomial order, in addition to the uncertainty determined by the
fitting routine.  If we add this 5\% uncertainty in quadrature to the
blue side measurement uncertainties, the agreement between the red and
blue sides appears more satisfactory, with $|\delta| \leq 1$ for 19 of
29 galaxies, or 65\% of the sample.  Thus, we conclude that the blue
side uncertainties are systematically too small and should be
increased by adding $0.05\sigma_B$ in quadrature to the uncertainties
listed in Table 3.  Increasing the error bars by this amount leads to
a satisfactory comparison between the red and blue measurements,
giving roughly the level of agreement that would be expected from
$1\sigma$ uncertainties in the case of Gaussian statistics.  It is
still somewhat surprising how large the disagreement is between the
blue and red data for a few of the galaxies, but in some cases (such
as NGC 4569) the discrepancy could be ascribed to a poor template
match due to the presence of a young starburst component.

\begin{figure}
\plotone{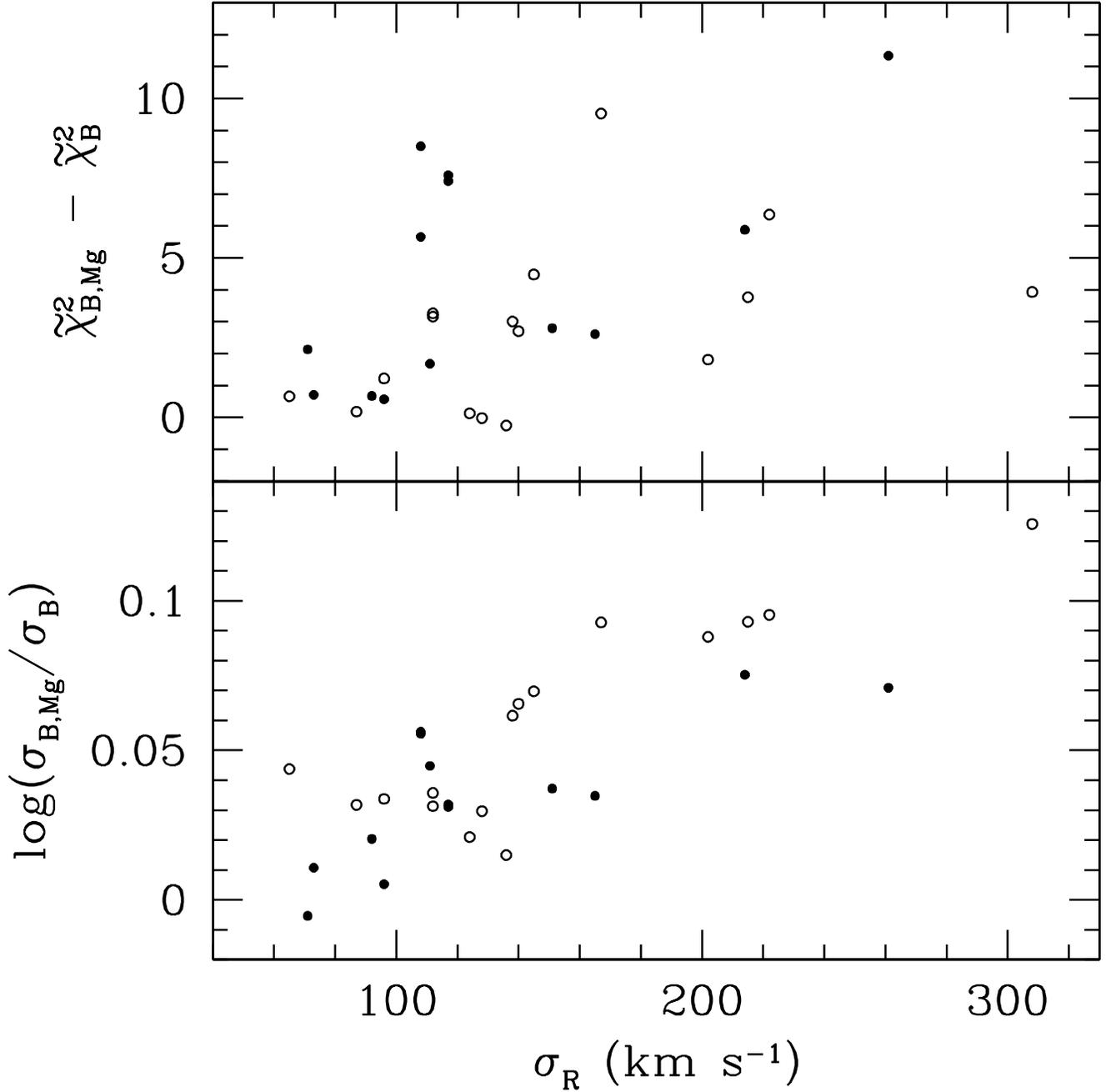}
\caption{Comparison of blue side measurements obtained by either
  excluding or including the \mgb\ lines from the fit.  \emph{Upper
  panel:} The difference in \chisq\ per degree of freedom for fits
  obtained either including ($\chisqdof_{B,Mg}$) or excluding
  ($\chisqdof_B$) the \mgb\ lines.  \emph{Lower panel:} Log of the
  ratio of velocity dispersions determined with or without \mgb\ in
  the fitting region.  Filled circles denote galaxies observed with the
  600 line grating on the blue side, and open circles denote galaxies
  observed with the 1200 line grating.
\label{mg}}
\end{figure}

Figure \ref{mg} demonstrates the results obtained if the \mgb\ lines
are included in the fitting region.  For comparison, we performed
measurements of our blue side data using the entire wavelength range
5040--5430 \AA, except for a small 20 \AA\ window centered on the
[\ion{N}{1}] emission line at 5200 \AA.  Including the Mg lines in the
fit nearly always leads to a significant increase in \chisqdof\
because of the poor match between the [Mg/Fe] ratios of the templates
and the galaxies.  The outcome is that, for all but one galaxy, the
velocity dispersions measured by including \mgb\ in the fitting region
are larger than those obtained with our default fitting region.  The
discrepancy increases systematically as a function of $\sigma$ due to
the correlation between [Mg/Fe] and $\sigma$, and is as bad as
25--30\% for galaxies with large $\sigma$.  The large increase in
\chisqdof\ clearly demonstrates that template stars of near-solar
metallicity should not be used to fit the Mg and Fe lines of galaxy
spectra simultaneously.  This template matching problem appears to
affect the measurements over the entire range in $\sigma$, not just
galaxies with large velocity dispersions.  Even for galaxies with
$\sigma\approx100$ \kms, the model fits are still severely degraded in
quality by including \mgb, and the velocity dispersions are affected
at the $\sim5-10\%$ level.

The difficulty in matching the \mgb\ line strength is a problem for
the direct-fitting method in particular, because the widths and the
depths of the absorption lines are coupled together in the calculation
of \chisq.  Methods for measurement of velocity dispersions that
operate in the Fourier domain may be less sensitive to variations in
line strength for individual lines.  The sensitivity of the Fourier
quotient method to metallicity variations in the \mgb\ region has been
examined by \citet{ll85}, who found that the derived dispersions have
a modest dependence on [Fe/H]; it would be worthwhile to perform
similar tests for various other measurement techniques, using galaxies
and template stars with a range of [Mg/Fe] ratios.

The [Mg/Fe] mismatch problem can be seen in some previous kinematic
studies, in cases where velocity dispersion measurements were
performed using a small wavelength region containing both \mgb\ and Fe
5270.  For example, some of the model fits shown by \citet{rw92} and
\citet{km93} appear to underpredict the strength of the \mgb\ lines
relative to Fe 5270.  In kinematic studies that attempt to derive the
shape of the LOSVD by methods operating in the pixel domain, it is
important to be aware of this [Mg/Fe] mismatch problem, because it
does affect the velocity dispersions (as shown in Figure \ref{mg}) and
there is the possibility that it could affect the shape of the derived
velocity profile as well.  As Figure \ref{bfits} demonstrates, this
problem can be largely avoided by shifting the fitting region redward
to cover the Fe 5270 and Fe 5335 blends, and excluding \mgb.

\subsection{Comparison with previous results}

\begin{figure}
\plotone{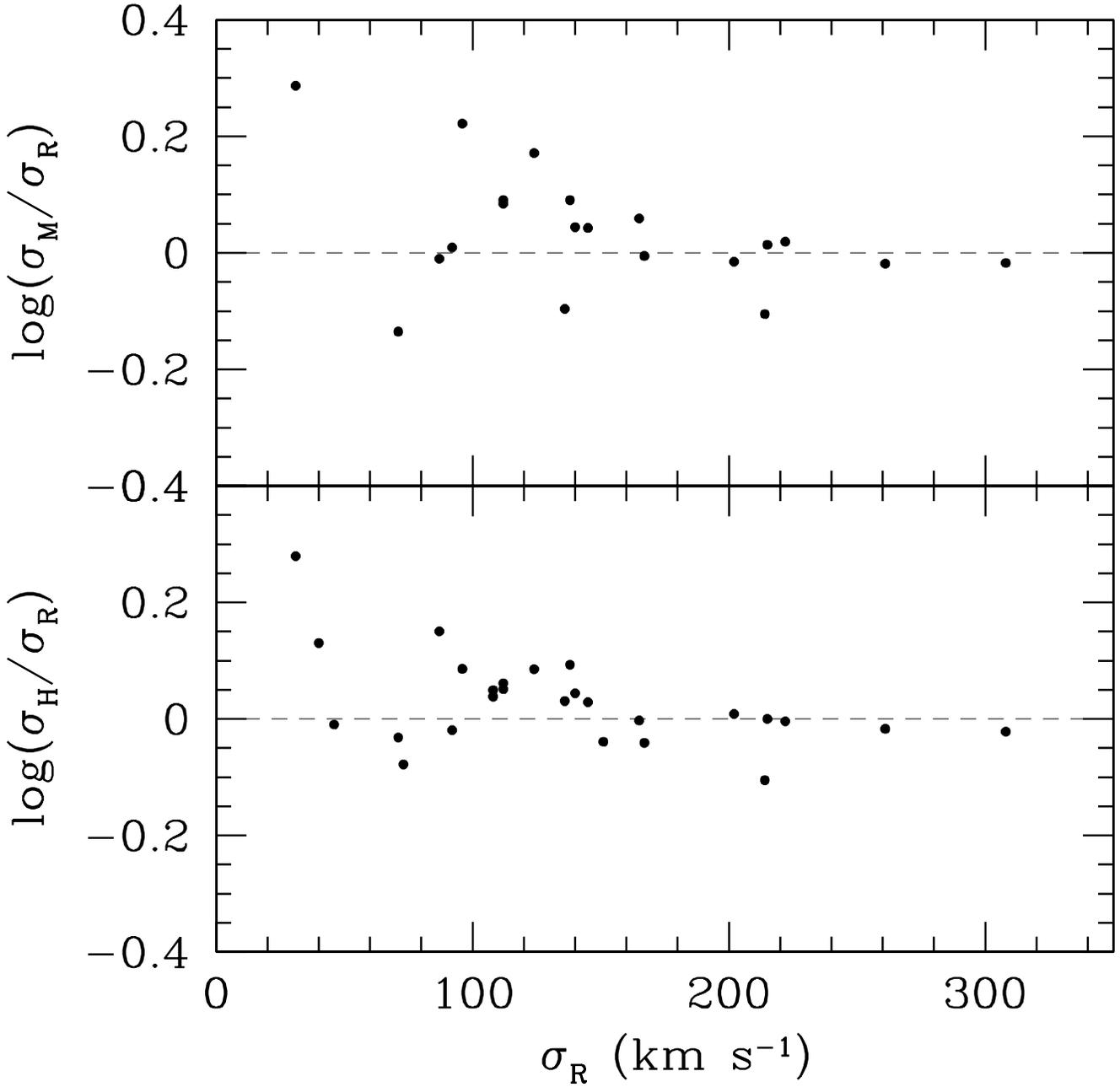}
\caption{Comparison of Ca triplet measurements of $\sigma$ with data
  collected from the literature by \citet{mce95} and by Hypercat.
  \emph{Upper panel:} Ratio of mean data listed by McElroy
  ($\sigma_M$) to our red side results.  \emph{Lower panel:}
  Ratio of mean velocity dispersion values listed by Hypercat
  ($\sigma_H$) to our red side results.
\label{litcompare}}
\end{figure}

Velocity dispersions have been reported previously for all but 6 of
the galaxies in our sample.  Two of the most commonly used references
for velocity dispersions measurements are the compilation by
\citet{mce95}, which lists averages of measurements from the
literature for each galaxy, and the online Hypercat database
\citep{pru98}, which lists all previous measurements of $\sigma$ and
also computes an average value for each galaxy.\footnote{The Hypercat
database is available at http://www-obs.univ-lyon1.fr/hypercat .}
Figure \ref{litcompare} shows a comparison of our results with the
average values reported by \citet{mce95} and by Hypercat.  The
measurements compiled by McElroy are corrected to a standard aperture
of $2\arcsec\times4\arcsec$, very close to the aperture size we used,
so the results should be directly comparable.  Hypercat also applies
correction factors to homogenize the measurements to a consistent
aperture size prior to computing the average values.

In general, the Hypercat mean results agree more closely with our red
side results than do the McElroy average values.  This is primarily
due to the fact that the Hypercat catalog is more up-to-date and
contains a larger number of measurements for some galaxies, so the
Hypercat averages are sometimes less affected by individual discrepant
measurements.  For the 20 galaxies in common between our sample,
\citet{mce95}, and Hypercat, the RMS deviation between the catalog
mean results and our red side measurements is 28 \kms\ for McElroy,
and 20 \kms\ for Hypercat.  Since most of the galaxies in our sample
are spirals, some portion of this scatter must be due to different
slit position angles, which would lead to different amounts of
rotational broadening in the extracted spectrum.

The averages computed by these catalogs often include quite discrepant
measurements taken from different literature sources.  The case of NGC
3627 serves as a useful example.  Hypercat lists two sources for
$\sigma$ that disagree by far more than their quoted uncertainties:
$117\pm9$ \kms\ \citep{hs98}, and $184\pm19$ \kms\ \citep{wks79}.  Our
result ($124 \pm 3$ \kms) agrees well with \citet{hs98} but is
significantly lower than that of \citet{wks79}.  Similarly, for NGC
4826, \citet{wks79} find $\sigma=160$ \kms\ while two other sources
listed in Hypercat give 90 and 113 \kms, closer to our red side
measurement of 96 \kms.  These examples demonstrate that the average
values listed by Hypercat, and by McElroy, should be viewed with a
great deal of caution, because the average dispersions for galaxies
with a small number of measurements can be badly influenced by a
single discrepant value.  There appears to be a systematic problem with
the measurements of \citet{wks79} in particular.  There are 8 galaxies
in common between our sample and \citet{wks79}, and in only one case
(NGC 4374) is their velocity dispersion smaller than our result.  On
average their measurements are larger than ours by 25\%.  The
sytematic offset of this one source contributes a nonnegligible amount
to the overall disagreement between our results and the average
literature results.

The worst apparent disagreement between our results and the mean
literature data is for NGC 1058, a late-type spiral galaxy with a
compact central stellar cluster for which we find $\sigma = 31 \pm 6$
\kms.  The mean velocity dispersion is given by \citet{mce95} as 60
\kms, and by Hypercat as 59 \kms.  However, according to Hypercat, the
sole previous measurement is from unpublished data of Whitmore \&
Rubin (1985), and according to Hypercat the original measurement was
actually an \emph{upper limit} of 60 \kms.  This example serves as a
reminder that when velocity dispersions for individual galaxies are
taken from the literature, it is probably safer to consult the
original sources than to rely on averaged results reported in these
compilations.

\section{Summary}

We have measured the central velocity dispersion within a
$2\arcsec\times3\farcs7$ aperture for 33 galaxies, using an extremely
simple fitting routine.  The measurements are performed by fitting
broadened stellar templates to the galaxy spectra, assuming a Gaussian
model for the LOSVD. Results obtained from the Ca triplet lines appear
to be quite robust and are not very sensitive to the choice of
template star or to the particular way in which the continuum shape of
the template is adjusted to match the galaxy; this is consistent with
the conclusions of \citet{dre84}.  The \mgb\ region is subject to
somewhat larger uncertainty, as it is a more complicated spectral
region and more susceptible to template mismatching.  We find that
after taking into account an estimated $\sim5\%$ uncertainty in the
blue side measurements due to the choice of a particular form for the
continuum normalization function, the blue and red measurements are in
agreement within their uncertainties for 65\% of the sample.

Template mismatch in the \mgb\ region can be reduced by avoiding the
use of both Mg and Fe lines simultaneously in measurement of $\sigma$,
because the [Mg/Fe] abundance ratio is correlated with $\sigma$.
Typical nearby K-giant stars used as velocity templates will not match
the [Mg/Fe] ratios of galaxies having large velocity dispersions, and
fits of broadened templates to galaxy spectra over the $5000-5500$
\AA\ region are often dramatically improved if the \mgb\ lines are
excluded from the fit.  This template matching problem should be
considered when performing more detailed analyses of galaxy
kinematics, such as measurements of black hole masses using
spatially-resolved spectra of the \mgb\ region from the \emph{Hubble
Space Telescope}.

\acknowledgments

We thank Todd Small for assistance during the June 2001 observing run,
Tom Matheson for providing software used in the data reduction, and
Pieter van Dokkum for helpful discussions.  Research by A.J.B. is
supported by NASA through Hubble Fellowship grant \#HST-HF-01134.01-A
awarded by the Space Telescope Science Institute, which is operated by
the Association of Universities for Research in Astronomy, Inc., for
NASA, under contract NAS 5-26555.  Research by W.L.W.S. is supported
by NSF grant AST-9900733.  This research has made use of the NASA/IPAC
Extragalactic Database (NED) which is operated by the Jet Propulsion
Laboratory, California Institute of Technology, under contract with
NASA.

\clearpage

\begin{center}
\begin{deluxetable}{lcccccc}
\label{observingruns}
\tablewidth{7in} 
\tablecaption{Observing Runs} 

\tablehead{\colhead{UT Date} & \colhead{Observatory} & \colhead{Slit}
& \colhead{Grating} & \colhead{$\lambda$} & \colhead{Plate Scale} &
\colhead{$\sigma$(instrumental)}  \\ & &
\colhead{(\arcsec)} & \colhead{(lines mm\per)} & \colhead{(\AA)} &
\colhead{(\AA\ pixel\per)} & \colhead{(\kms)\tablenotemark{a}}  }

\startdata 

2001 Jun 21--25 & Palomar & 2 & Blue: 600 & $4210-5950$ & 1.72 & 115 \\
 & & & Red: 1200 & $8400-9070$ & 0.63 & 25 \\
2002 Jan 24--26 & Palomar & 2 & Blue: 1200 & $4940-5820$ & 0.88 & 60 \\
 & & & Red: 1200 & $8400-9070$ & 0.63 & 25 \\
\enddata

\tablenotetext{a}{Instrumental dispersion for a source uniformly
  filling the slit, measured from the widths of comparison lamp lines
  near 5200 \AA\ and 8500 \AA.}
\end{deluxetable}
\end{center}
\normalsize

\begin{center}
\begin{deluxetable}{lcc|lcc}
\label{templatestars}
\tablewidth{4.5in} 
\tablecaption{Template Stars} 

\tablehead{\multicolumn{3}{c}{June 2001} & \multicolumn{3}{c}{January
    2002} \\ \colhead{Star} & \colhead{Type} & \colhead{[Fe/H]} &
    \colhead{Star} & \colhead{Type} & \colhead{[Fe/H]} }

\startdata

HD 121146 & K2 IV     &  $-$0.202 & HD 12929 & K2 III & 0.036 \\
HD 125560 & K3 III    &  0.133  & HD 19476 & K0 III & 0.101 \\
HD 129312 & G7 III    &  $-$0.097 & HD 20893 & K3 III & $-$0.001 \\
HD 136028 & K5 III    &  $-$0.110 & HD 49293 & K0 IIIa & 0.018 \\
HD 188056 & K3 III    &  0.088  & HD 51440 & K2 III & $-$0.565 \\
HD 199580 & K0 III-IV &  $-$0.17  & HD 58207 & G9 IIIb & $-$0.127 \\
HD 203344 & K1 III    &  $-$0.140 & HD 69267  & K4 III & $-$0.130 \\
HD 221148 & K3 III    &  0.133  & HD 125560 & K3 III & 0.133 \\

\enddata
\tablecomments{[Fe/H] values are from the compilation by \citet{tay99}.}
\end{deluxetable}
\end{center}

\begin{center}
\begin{deluxetable}{lcccccc}
\label{datatable}
\tablewidth{5in} 
\tablecaption{Velocity Dispersions} 

\tablehead{\colhead{Galaxy} & \colhead{Type} & \colhead{Slit P.A.} & \colhead{Exposure} &
\colhead{Blue Grating} & \colhead{$\sigma_B$} &
 \colhead{$\sigma_R$} \\ & & \colhead{($\arcdeg$)} & \colhead{(s)}
& (lines mm\per) & \colhead{(\kms)} & \colhead{(\kms)} }

\startdata 
Arp 102B & E0      & 298 & 4800 & 600  & \nodata      & $188 \pm 8$ \\
NGC 404  & SA0     & 80  & 1200 & 1200 & \nodata      & $40 \pm 3$  \\
NGC 660  & SBa pec & 46  & 1800 & 1200 & $98 \pm  13$ & $128 \pm 6$ \\ 
NGC 1052 & E4      & 30  & 1800 & 1200 & $215 \pm  7$ & $215 \pm 4$ \\
NGC 1058 & SAc     & 90  & 3600 & 1200 & \nodata      & $31 \pm 6$  \\
NGC 2787 & SB0     & 60  & 2400 & 1200 & $218 \pm  7$ & $202 \pm 5$ \\
NGC 2841 & SAb     & 70  & 2400 & 1200 & $229 \pm  5$ & $222 \pm 4$ \\
NGC 3486 & SABc    & 105 & 1200 & 1200 & $64 \pm 4$   & $65 \pm 3$  \\
NGC 3489 & SAB0    & 160 & 900  & 1200 & $100 \pm 4$  & $112 \pm 3$ \\
NGC 3623 & SABa    & 90  & 1200 & 1200 & $149 \pm 4$  & $138 \pm 3$ \\
NGC 3627 & SABb    & 160 & 900  & 1200 & $123 \pm 6$  & $124 \pm 3$ \\
NGC 3675 & SAb     & 86  & 2400 & 600  & $120 \pm 11$ & $108 \pm 4$ \\
NGC 3982 & SABb    & 110 & 2400 & 600  & $81 \pm 13$  & $73 \pm 4$  \\
NGC 4150 & SA0     & 90  & 900  & 1200 & $74 \pm  6$  & $87 \pm 3$  \\
NGC 4203 & SAB0    & 95  & 1500 & 1200 & $175 \pm 4$  & $167 \pm 3$ \\
NGC 4278 & E1-2    & 70  & 2400 & 600  & $251 \pm 13$ & $261 \pm 8$ \\
NGC 4314 & SBa     & 72  & 2400 & 600  & $119 \pm  9$ & $117 \pm 4$ \\
NGC 4321 & SABbc   & 54  & 2100 & 600  & $101 \pm 12$ & $92 \pm 4$  \\
NGC 4374 & E1      & 150 & 1500 & 1200 & $302 \pm 7$  & $308 \pm 7$ \\
NGC 4414 & SAc     &  73 & 2400 & 600  & $128 \pm  9$ & $117 \pm 4$ \\
NGC 4494 & E1-2    & 160 & 1200 & 1200 & $155 \pm 4$  & $145 \pm 3$ \\
NGC 4569 & SABab   & 150 & 900  & 1200 & $114 \pm 7$  & $136 \pm 3$ \\
NGC 4579 & SABb    & 55  & 1800 & 600  & $166 \pm 8$  & $165 \pm 4$ \\
NGC 4639 & SABbc   & 54  & 1800 & 600  & $71 \pm 15$  & $96 \pm 4$  \\
NGC 4725 & SABab   & 124 & 900  & 1200 & $136 \pm 4$  & $140 \pm 3$ \\
NGC 4736 & SAab    & 124 & 900  & 1200 & $109 \pm 4$  & $112 \pm 3$ \\
NGC 4800 & SAb     & 99  & 3600 & 600  & $ 90 \pm 12$ & $111 \pm 4$ \\
NGC 4826 & SAab    & 124 & 900  & 1200 & $99 \pm 4$   & $96 \pm 3$  \\
NGC 5033 & SAc     & 73  & 2400 & 600  & $116 \pm 11$ & $151 \pm 4$ \\
NGC 5055 & SAbc    &  83 & 2400 & 600  & $111 \pm 11$ & $108 \pm 4$ \\  
NGC 5273 & SA0     & 80  & 3600 & 600  & $73 \pm 14$  & $71 \pm 4$  \\
NGC 6500 & SAab    & 10  & 3600 & 600  & $203 \pm 11$ & $214 \pm 6$ \\
NGC 6503 & SAcd    & 1   & 3600 & 600  & \nodata      & $46 \pm 3$  \\

\enddata

\tablecomments{Morphological types are from NED.  We consider the
  velocity dispersions measured from the red side to be more accurate
  than the blue side measurements, for reasons described in the text.}

\end{deluxetable}
\end{center}

\end{document}